\documentclass[10pt,conference]{IEEEtran}
\usepackage{graphicx,times,amsmath,amsfonts}
\usepackage{listings}
\begin{document}    

%

%
\author{\IEEEauthorblockN{Shahriar Shahabuddin, Janne Janhunen, and Markku Juntti}\\
\IEEEauthorblockA{Department of Communications Engineering and Centre for Wireless Communications, University of Oulu, Finland.}\\
\IEEEauthorblockA{\{sshahabu, janne.janhunen, markku.juntti\}@ee.oulu.fi}}
%

\newcommand{\MYheader}{\smash{\scriptsize
\hfil\parbox[t][\height][t]{\textwidth}{\centering {\normalsize
Reference Matlab code is available at $ sites.google.com/site/shahriarshahabuddin/matlab\_simulator $}}\hfil\hbox{}}}
\makeatletter
\def\ps@headings{%
\def\@oddhead{\MYheader}
\def\@evenhead{\MYheader}
\def\@oddfoot{  }%
\def\@evenfoot{  }}
\def\ps@IEEEtitlepagestyle{%
\def\@oddhead{\MYheader}%
\def\@evenhead{\MYheader}%
\def\@oddfoot{   }%
\def\@evenfoot{   }}
\makeatother
\pagestyle{headings}
\addtolength{\footskip}{0\baselineskip}
\addtolength{\textheight}{-0.1\baselineskip}

\title{Design of a Transport Triggered Architecture Processor for Flexible Iterative Turbo Decoder}

\maketitle
\begin{abstract}
\boldmath{In order to meet the requirement of high data rates for the next generation wireless systems, the efficient implementation of receiver algorithms is essential. On the other hand, the rapid development of technology motivates the investigation of programmable implementations. This paper summarizes the design of a programmable turbo decoder as an application-specific instruction-set processor (ASIP) using Transport Triggered Architecture (TTA). The processor architecture is designed in such manner that it can be programmed to support other receiver algorithms, for example, decoding based on the Viterbi algorithm. Different suboptimal maximum $\it{a}$ $\it{posteriori}$ (MAP) algorithms are used and compared to one another for the soft-input soft-output (SISO) component decoders in a single TTA processor. The max-log-MAP algorithm outperforms the other suboptimal algorithms in terms of latency. The design enables the designer to change the suboptimal algorithms according to the bit error rate (BER) performance requirement. Unlike many other programmable turbo decoder implementations, quadratic polynomial permutation (QPP) interleaver is used in this work for contention-free memory access and to make the processor 3GPP LTE compliant. Several optimization techniques to enable real time processing on programmable platforms are introduced. Using our method, with a single iteration 31.32 Mbps throughput is achieved for the max-log-MAP algorithm for a clock frequency of 200 MHz. }
\end{abstract}
\section{Introduction}\label{1}

The turbo coding scheme [1] has been adopted for the air interface standard called Long Term Evolution (LTE), that has been defined by the 3rd Generation Partnership Project (3GPP) [2]. The decoding algorithm for the component decoder is not specified by 3GPP. Different suboptimal algorithms have been used for the turbo decoding and new solutions with less complexity are still being invented quite frequently. A programmable implementation simplifies the necessary support for different suboptimal algorithms of component decoders and different interleaving methods.

The turbo decoding algorithm is still one of the most computationally intensive parts of the wireless receiver. The software implementations provide the required flexibility to support multistandard solutions, but requires a careful design to achieve the target throughput. On the other hand, the hardware implementations provide high throughput, but the development time is not as rapid as processor based implementations. Programmable accelerators, which enable software-hardware co-design method might be an attractive solution to overcome these bottlenecks. The design of software and hardware together to grind out the best performance and ensure programmability is not a straightforward task. The designer needs a very efficient tool, which can be used to design the processor easily for a particular application.

In this paper, we propose a design of a processor based on the Transport Trigger Architecture (TTA) for a flexible turbo decoder. TTA is a very good processor template for a programmable application-specific instruction-set processor (ASIP). The TTA based codesign environment (TCE) tool enables the designer to write the application with high level language and design the target processor in a graphical user interface at the same time [3].

The turbo decoder based on the transport triggred architecture was implemented by Salmela $\it{et}$ $\it{al}$. [4]. The throughput of the processor was comparable to those of the turbo decoders based on a pure hardware design, but was less flexible in comparison with the design presented in this paper. The interleaver of the processor was not presented according to the 3GPP LTE standard. 

The focus of the turbo decoder processor design in this paper is flexibility. The quadratic permutation polynomial (QPP)
 interleaving pattern have been used for the interleaver block. 
The comparison with the other suboptimal algorithms like log-MAP, constant-log-MAP and linear-log-MAP has also been presented. 

The rest of the paper is organized in the following way: In Section \ref{2}, an overview of the turbo decoder has been presented. In Section \ref{3}, the simplification techniques of the turbo decoding algorithm and its implementation in high level language is presented. The processor design has been presented in Section \ref{4}. In Section \ref{5}, we will present the throughput results of the design. The conclusion is given in Section \ref{6}.

\section{Turbo Decoder} \label{2}
\subsection{Turbo Decoder Structure}
The turbo decoder consists of two soft-input soft-output (SISO) decoders, with interleavers and de-interleavers between them as shown in Fig. 1. The inputs of the turbo decoder come from the soft demodulator, which produces the log-likelihood ratios (LLR) for the systematic bits and parity bits. The LLRs  of the systematic bit, LuI and first parity bits, LcI1 goes to the first SISO decoder.  The SISO decoder produces soft outputs based on  these LLRs. These soft outputs are used in the second SISO decoder as the additional information. The inputs of the second SISO decoder are the LLRs coming from the systematic bits, second parity bits denoted by LcI2 and output of the first SISO decoder. The LLRs of the systematic bits are scrambled this time with the same interleaving pattern used at the encoder. Similarly, the soft outputs coming from the first SISO decoder are scrambled also with the same interleaving pattern, which are used as $a$ $priori$ values for the second SISO decoder. 

\begin{figure}[h]
\includegraphics[keepaspectratio,width=1\columnwidth]{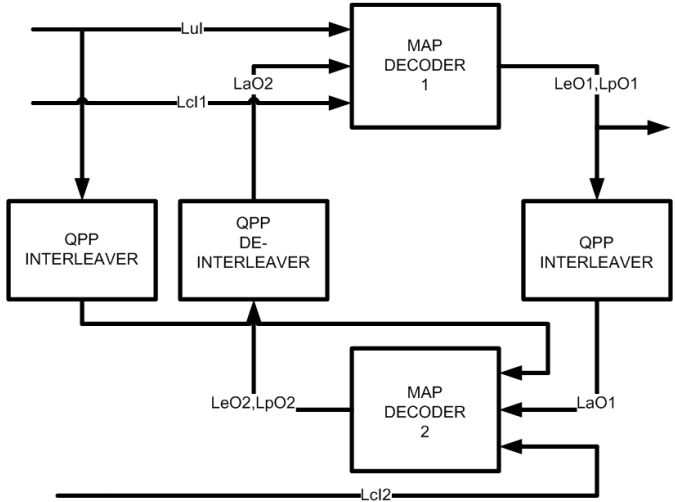}
\caption{Block diagram of the turbo decoder.}
\label{fig:}
\end{figure}
\vspace{-.5cm}
The heart of the turbo coding is the iterative decoding procedure. The output of the second SISO decoder does not produce the hard outputs immediately, but the soft output is used again in the first SISO decoder for more accurate approximation. The process continues in a similar fashion in an iterative manner.
A single iteration by both the first and the second SISO decoder is referred to as a full iteration. On the other hand, the operation performed by a single SISO decoder can be referred to as a half iteration. At the beginning of the first iteration, the $a$ $priori$ values are set at zero. Six to eight full iterations are used to achieve sufficient performance [1].

\subsection{MAP algorithm for Component Decoder}
The MAP algorithm for the component decoder applied here has been proposed by Benedetto $\it{et}$ $\it{al}$. [5]. The algorithm can be stated like:

1.  Initialize the values of the forward state metric as $\alpha_{0}(s)=0$ if $s=S_{0}$ and $\alpha_{0}(s)=-\infty $ otherwise.

2. Calculate all the forward state metric of the same window through the forward recursion according to 
\begin{equation}
\begin{split}
\alpha_{k}(s)=\hspace{.1cm}&\text{max}^*(\alpha_{k-1}[s^{S}(e)]+u(e)LuI[k-1]\\
&+c_1(e)LcI1[k-1]+c_2(e)LcI2[k-1]).
\end{split}
\end{equation}

3. Initialize the values of the backward state metric as $\beta_{n}(s)=0$ if $s=S_{n}$ and $\beta_{n}(s)=-\infty $ otherwise.

4. Calculate all the backward state metric of the same window through the backward recursion as
\begin{equation}
\begin{split}
\beta_{k}(s)=\hspace{.1cm}&\text{max}^*(\beta_{k+1}[s^{E}(e)]+u(e)LuI[k+1]\\
&+c_1(e)LcI1[k+1]+c_2(e)LcI2[k+1]).
\end{split}
\end{equation}

5. The LLR values for the information and both parity bits can be calculated as following:
\begin{equation}
\begin{split}
LLR(.;O)=\hspace{.1cm}&\text{max}^*(\alpha_{k-1}[s^{S}(e)]+c_1(e)LcI1[k-1]\\
&+c_2(e)LcI2[k+1]+\beta_{k+1}[s^{E}(e)]).
\end{split}
\end{equation}

The forward metric and backward metrics increase in each step and that is why the forward and backward metrics need to be normalized to avoid memory overflow. 

The decoding is done in smaller windows so that the decoding process can be done in parallel and the decoder does not have to wait for the whole block to arrive before starting the decoding process. This windowing is sometimes referred to as a sliding window method.

Based on the definition of the $\text{max}^*$ function, the suboptimal algorithms can be changed for the MAP algorithm. As for an example, for constant-log-MAP algorithm it can be expressed as 
\begin{equation}
\text{max}^*(x,y)=\text{max}(x,y)+\begin{cases}0 & if|y-x|>T\\C & if|y-x|\leq T.\end{cases}
\end{equation}
We considered four suboptimal MAP algorithms, which are Max-log-MAP, constant log MAP, linear-log-MAP and an approximation of log-MAP algorithm. A more detailed overview of these algorithms can be found in [6], [7] and [8]. 


\subsection{QPP Interleaver}
The QPP interleaver has been adopted for the 3GPP LTE standard [3]. Unlike the earlier 3G interleavers, the QPP interleaver is based on algebric constructions. The QPP interleaver can be expressed by a simple mathematical formula. The relationship between the output index $x$ and the input index $f(x)$ can be derived from this formula given below
\begin{equation}
    f(x)=(f_2x^2+f_1x)\text{mod}(N).
\end{equation}
The values of the parameters $f_1$ and $f_2$ of (5) are integers and depend on the block size $\it{N}$. For a different $\it{N}$, a different set of parameters have been defined in [9]. The block sizes are divisible by 4 and 8. The value of $f_1$ is an odd number while the value of $f_2$ is an even number. Several algebric properties of QPP interleaver have been listed in [10].

\section{Design in High Level Language} \label{3}
The MAP algorithm described above is complex and needs to be simplified. The repeated calculations should be avoided, which in turn reduces the latency of the processor. The reduced calculations used in this work will be described in the following section. The reduced design has been tested in a link level simulator using Matlab. The Matlab design is converted to C language in such a way so that it can be efficiently implemented in the TCE tool.
\begin{figure}[h]
\centering
\includegraphics[keepaspectratio,width=.7\columnwidth]{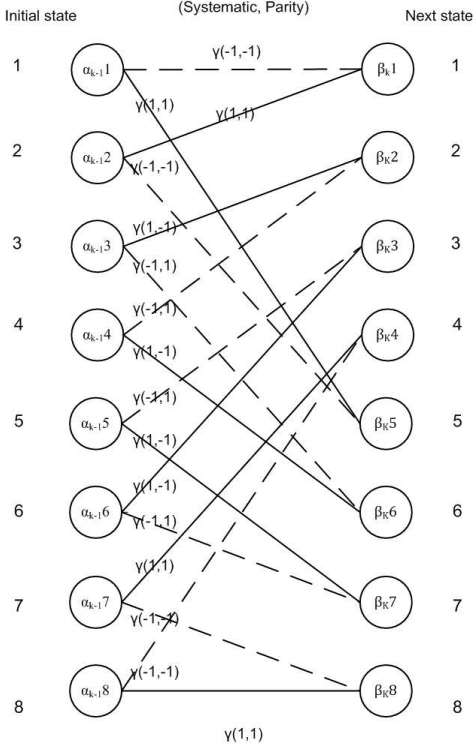}
\caption{Trellis of 3GPP turbo code.}
\label{fig:}
\end{figure}

There are 16 branch metric computations between two states for forward metric, backward metric and LLR calculations. From the trellis structure of the 3GPP turbo code it can be seen that four calculations of branch metric are being repeated to result in total sixteen calculations. Instead of multiplying with 1 and $-1$, the branch metrics can be calculated directly based on these four calculations,
\begin{equation}
\begin{split}
    \gamma_1 = LuI+LcI1+LcI2\\
	 \gamma_2 = -LuI-LcI1+LcI2\\
	 \gamma_3 = LuI+LcI1-LcI2\\
    \gamma_4 = -LuI-LcI1-LcI2,
\end{split}
\end{equation}
where $\gamma_4$ can be represented as $-\gamma_1$ and $\gamma_3$ can be represented as $-\gamma_2$. For every forward metric, the backward metric and the LLR, calculations of two branch is sufficient. As an example, the backward metric calculations are shown as, 
\begin{equation}
\begin{split}
\beta_1(k) = \text{max}^*(\beta_1(k+1)-\gamma_1,\beta_5(k+1)+\gamma_1)\\
\beta_2(k) = \text{max}^*(\beta_1(k+1)+\gamma_1,\beta_5(k+1)-\gamma_1)\\
\beta_3(k) = \text{max}^*(\beta_6(k+1)+\gamma_2,\beta_2(k+1)-\gamma_2)\\
\beta_4(k) = \text{max}^*(\beta_6(k+1)-\gamma_1,\beta_2(k+1)+\gamma_1)\\
\beta_5(k) = \text{max}^*(\beta_1(k+1)-\gamma_1,\beta_5(k+1)+\gamma_1)\\
\beta_1(k) = \text{max}^*(\beta_1(k+1)-\gamma_1,\beta_5(k+1)+\gamma_1)\\
\beta_1(k) = \text{max}^*(\beta_1(k+1)-\gamma_1,\beta_5(k+1)+\gamma_1)\\
\beta_1(k) = \text{max}^*(\beta_1(k+1)-\gamma_1,\beta_5(k+1)+\gamma_1).
\end{split}
\end{equation}
\vspace{-.2cm}

A similar idea has been presented in [11] where a simplified version of max-log-MAP has been proposed. 
The forward metrics have been calculated first in this design by following the simplified equation. The normalization technique is done following the technique suggested by Valenti $\it{et}$ $\it{al}$. [12]. Typically, the normalization is done in every step by subtracting the minimum values of forward or backward metrics of the same column from every forward or backward metric values. They  suggested subtracting all the values from the first values of the columns. Thus, there is no need to store the first row and this constitutes 12.5 percent savings in memory compared with the other normalization methods.

Only the forward metrics are saved for the calculation of the LLRs. The backward metrics and LLR calculations are done together. As soon as the backward metric calculations of a single column are finished, the LLRs are calculated in parallel with the help of the stored forward metric at the same time. So there is no need to save the backward matrics [12].

Originally, $\alpha$ and $\beta$ are two-dimensional matrices of size $6144\times8$ as the input block of 6144 information bits have been taken into consideration. The processor needs to do more calculation for accessing these two dimensional matrices. Instead of using a single $6144\times 8$ matrix, seven vectors of 6144 elements have been used. As mentioned above, the eighth vector is not needed due to the normalization strategy adopted in the used method.

The turbo decoder description in C needs to be efficient to fulfill the latency requirement. A lot of data dependencies can be seen from the turbo decoder algorithm. So, writing efficient code is needed for the utilization of computing resources available.



Assuming thatenough register files are available, efficient code is written to reduce memory accesses by holding the values inside the register files. 
It is also possible to eliminate the need for accessing the forward metric vectors constantly. Instead, the temporary variables are used to hold the values inside the registers. The code for the backward metric and the LLR calculations have been written in the same manner. The calculations are done is several smaller loops to follow the sliding window technique and to enable parallel calculations. 
\begin{figure*}[!t]
\centering
\includegraphics[keepaspectratio,width=2\columnwidth]{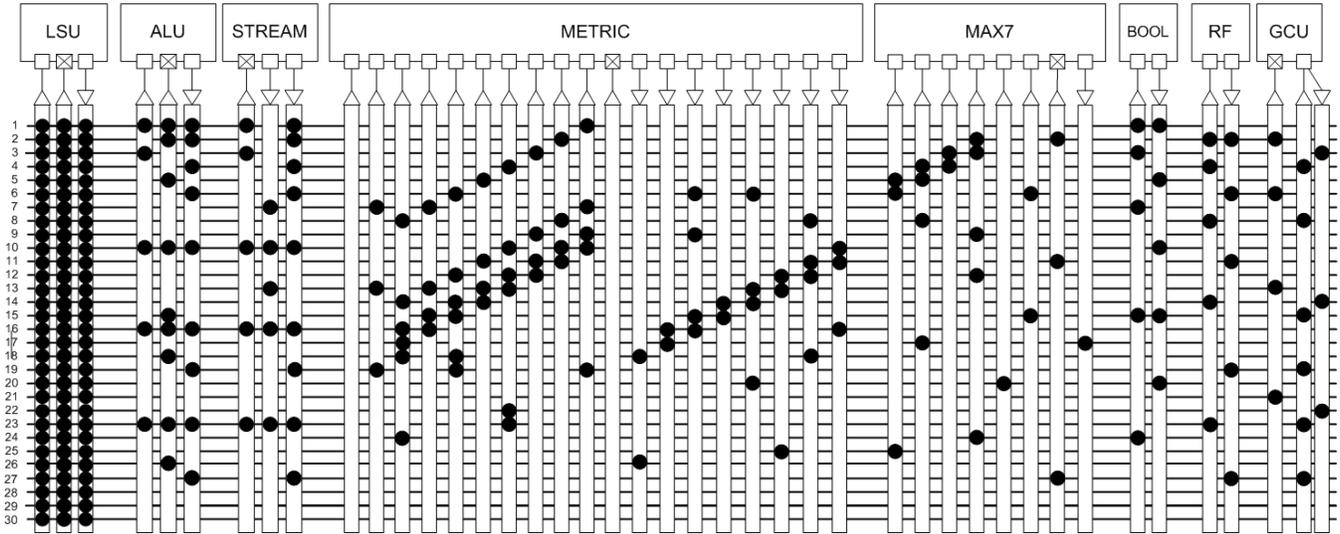}
\caption{Implemented processor with reduced number of functional units.}
\label{fig:}
\vspace*{4pt}
\end{figure*}

A look-up table, holding the values of the $f_1$ and $f_2$ parameters, is used to implement the QPP interleaver. The rest of the calculations are quite straightforward.

\section{Transport Triggered Architecture Processor} \label{4}
\subsection{General Functional Units}
A part of the processor designed for the turbo decoder is illustrated in Fig. 3. For readability, the whole processor figure is not given. The blocks on the upper part of the figure represent the functional units and register files of the processor. The black horizontal straight lines represent the buses of the processor. The vertical rectangular blocks represent the sockets. The connection between functional units and buses is illustrated by black spots in the sockets.


The fixed point processor includes load/store unit (LSU), arithmetic logic unit (ALU), global control unit (GCU) and register files. Based on the resource requirements in high level language, more functional units and register files have been added. 

The LLR inputs have been read from a first-in-first-out (FIFO) memory buffer by using the functional unit called STREAM. The STREAM units can read every input sample in one clock cycle. Eight STREAM units have been used to get the input LLRs simultaneously. The STREAM units are used to implement the sliding window technique that helps to decode the input block in smaller parts parallely. One STREAM unit has been used to write the output LLRs in the memory buffer. 

Three LSU units have been used to support the memory accesses. The LSU units are used to read and write memory. The memory can be read in three clock cycles and can be written in a single cycle. 

The ALU unit has been used to perform the basic arithmetic operations like addition, subtraction etc. Operations like shifting right or left are also included in ALU.

\subsection{METRIC Special Function Unit}
Like the Viterbi decoder, the turbo decoder needs numerous add-compare-select operations. One way to design the processor for these algorithms is to use the appropriate number of adder and maximum selection units. Another way is to design a special function unit to calculate all the necessary next state metrics based on the earlier state metrics and branch metrics. 

The latter way is followed in our design and a special function unit named METRIC has been designed with twelve inputs and eight outputs. Eight of these inputs correspond to the forward metrics in case of forward metric calculations and backward metrics in case of backward metric calculations. Three of these inputs correspond to the $a$ $priori$ LLR, the LLR of systematic bits and the LLR of parity bits, respectively. One input is used to select different suboptimal algorithms and this input is named $mode$ in this paper. 

The forward and backward metrics calculations are not the same. The inputs and outputs of the unit need to be selected carefully because the same unit has been used for both forward and backward metric computations. Using the same unit for forward and backward metric computations is possible because of the trellis structure of a 3GPP turbo code.

The design of the Metric unit for one butterfly pair in the trellis is shown in Fig. 4. The values of branch metrics to calculate $\alpha_2(k)$ and $\beta_3(k)$ are the same. Likewise, the values of branch metrics to calculate $\alpha_6(k)$ and $\beta_4(k)$ are also the same. It is possible to calculate these forward and backward metrics by changing the input values of the earlier state metrics. This is true for all the rest of the butterfly pairs in the trellis. The METRIC unit uses this technique of changing the orders of the input metrics for the backward metrics calculation, but using the same function unit.

The METRIC unit supports four operations. These four operations are branch metric calculations for max-log-MAP, linear-log-MAP, constant-log-MAP and log-MAP respectively. The latency of these operations are different because these operations use different codes for different algorithms.

\begin{figure}[h]
  \centering
\includegraphics[keepaspectratio,width=3.5in]{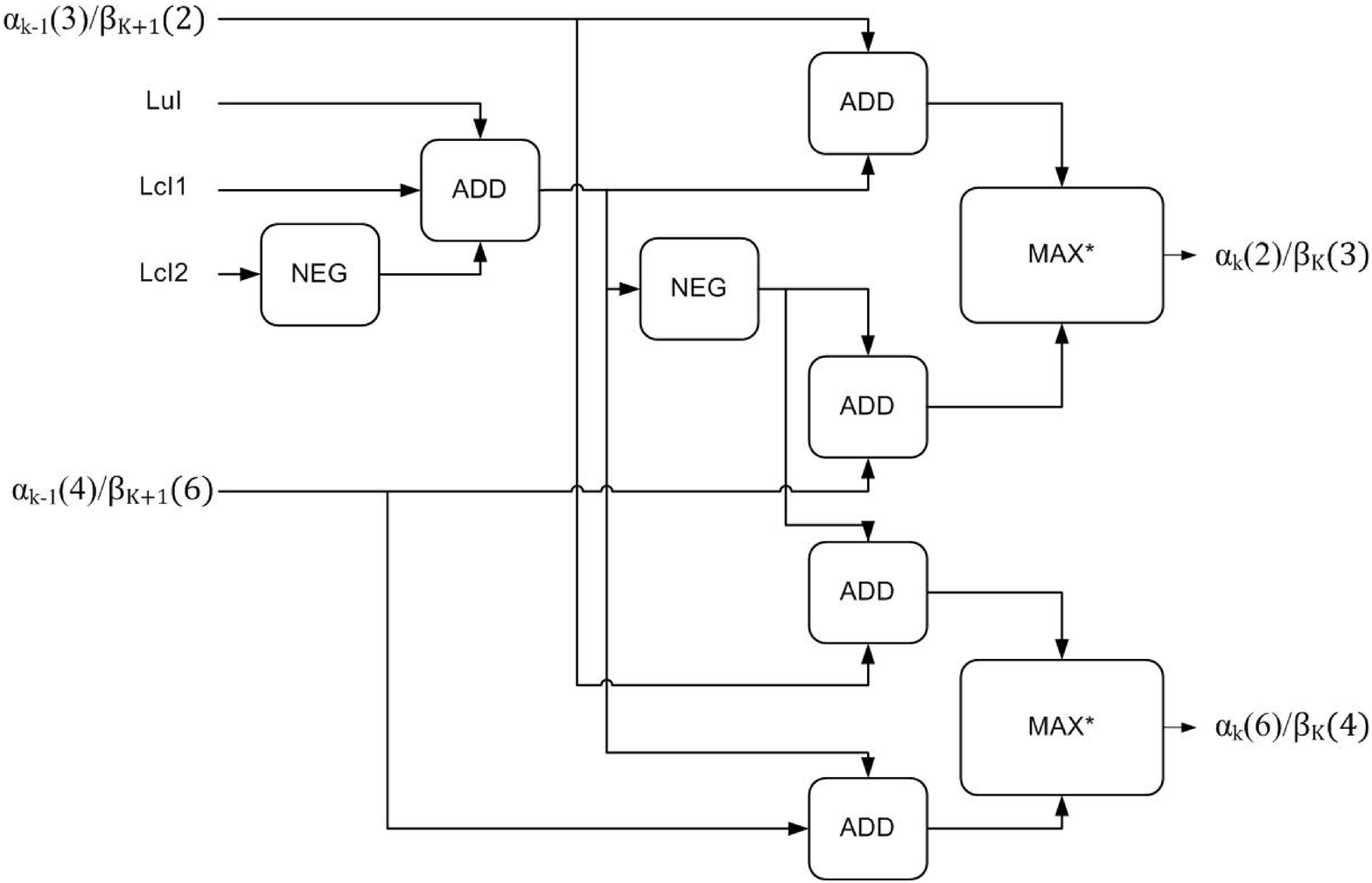}
  \caption{METRIC unit for a single butterfly pair.}
\end{figure}

\subsection{MAX7 Special Function Unit}
Another special function unit named MAX7 has been designed, which is used to calculate the output LLRs after the forward and backward metrics computation. This unit also has four options for four suboptimal algorithms.  For the max-log-MAP $\it{mode}$, the MAX7 unit takes seven inputs and finds the maximum of these inputs. The LLR calculation uses the maximum value of seven computations in this case. The addition of the forward metric, branch metric and next state backward metric represent these computations. It is possible to reduce one computation because of the normalization method chosen in this design. We have given the added values of these seven computations as inputs to this unit and used the output as the LLR. The latency of this unit is different for four different algorithms.
\subsection{Registers}
Several register files have been used to save the intermediate results. In terms of the power consumption, registers can be more expensive than memory, but to meet the latency requirement register files are needed. A single Boolean register file has been included in the processor design.

\section{Results and Discussion} \label{5}
The designed processor takes 39,226 clock cycles to process three blocks of 6,144 samples for a full iteration for the max-log-MAP mode. According to the 3GPP interleaver specifications, the size of the block could be up to of 6,144 samples and that is the reason the size of the input block is chosen as 6,144.

When we assume a clock of 200 MHz, then the throughput of one iteration for the max-log-MAP could be calculated as,
\begin{align}\nonumber
\text{Throughput} = &\frac{6,144\hspace{.1cm} \text{bits}\times  200 \hspace{.1cm}\text{MHz}}{39,226\hspace{.1cm} \text{clock cycles}}\\
&= 31.32\hspace{.1cm}\text{Mbps}.\nonumber
\end{align}

The clock cycle needed for a single trellis stage of the max-log-MAP can be calculated as,
\begin{align}\nonumber
\text{Cycle}_{Stage} = &\frac{39,226}{2 \times 6,144}\\
&= 3.19\hspace{.1cm}\text{cycles}.\nonumber
\end{align}

\begin{table}[h]
\centering
\caption{Number of clock cycles for a single iteration}
\label{tab1}
\begin{tabular}{|c|c|c|}
\hline
Modes & Algorithm & Clock Cycle\\
\hline
1 & max-log-MAP & 39,226\\
\hline
2 & linear-log-MAP & 103,436\\
\hline
3 & constant-log-MAP & 184,454\\
\hline
4 & log-MAP & 834,253\\
\hline
\end{tabular}
\end{table} 

It can be seen from Table \ref{tab1} that the processor takes more clock cycles if max-log-MAP algorithm is not used. The reason lies in the fact that the constant-log-MAP, linear-log-MAP and log-MAP algorithms need to invoke some conditional statements resulting in branches in execution. Thus, the latency increases compared to the max-log-MAP algorithm.
A less complicated approximation of log-MAP has been used in this design. However, the correction term of the log-MAP increases the latency significantly compared to the other three suboptimal algorithms even after the simplification. Linear-log-MAP is slower than the constant-log-MAP because of some multiplication operations are needed. The log-MAP algorithm provides better BER performance than the other algorithms. This log-MAP can be used in cases when the latency requirements are not strict and a high BER performance is needed.  

The buses of the processor are perfectly utilized to achieve the best possible result due to the perfect scheduling. The number of some of the operations during the algorithm execution has been summarized in Table \ref{2}.

\begin{table}[h]
\centering
\caption{Number of operations}
\label{tab2}
\begin{tabular}{|c|c|}
\hline

Operation & \# of OPS\\
\hline
ADD & 339,009\\
\hline
SUB & 98,304\\
\hline
MUL & 45,234\\
\hline
MAX7 & 24,567\\
\hline
BRANCH & 24,576\\
\hline
STREAM & 24,556\\
\hline
\end{tabular}
\end{table} 

The number of addition operations does not only represent the addition for the algorithm, but for several other purposes like loop indexing for the code. The subtraction operations are due to the normalization and the multiplication is used for the correction term calculation of linear-log-MAP and for the QPP interleaving sequence generation. 

The throughput can be increased using dedicated special function units working as accelerators for the processor. On the other hand, the more special function units are used, the less flexible the processor becomes. As an example, a special function unit dedicated to calculate the backward metric and the LLRs could be designed to reduce latency. However, it might be useless for some other functionalities. A common special function unit to calculate both forward and backward metric could be better utilized in this case.

The TTA processor designed here will work for the Viterbi decoding with reasonable throughput. The BRANCH unit for the max-log-MAP mode is able to calculate the add-select-compare operations needed for the Viterbi decoding.

A comparison with different other programmable implementations of turbo decoder has been presented in Table \ref{tab3}. Our proposed processor provides very good throughput compared to most of the programmable implementations. The processor throughput is lower than the TTA processor of [4] for the same clock frequency of 200 MHz, because of the complex structure of QPP interleaver and de-interleaver.
\begin{table}[h]
\centering
\caption{Programmable processors}
\label{tab3}
\begin{tabular}{|c|c|c|c|}
\hline

Reference & Architecture & Algorithm & Throughput\\
\hline
[13] & Motorola & max-log-MAP & 243 Kbps\\
\hline
[12] & Intel Pentium  & max-log-MAP & 366 Kbps\\
\hline
[14] & TMS320C6201 DSP & max-log-MAP & 2 Mbps\\
\hline
[15] & VLIW ASIP & max-log-MAP & 5 Mbps\\
\hline
[4] & TTA proc. for UMTS & max-log-MAP & 14.1 Mbps\\
\hline
Proposed & TTA proc. & max-log-MAP & 31.21 Mbps\\
\hline
[4] & TTA proc. & max-log-MAP & 98 Mbps\\
\hline
\end{tabular}
\end{table} 
  
\section{Conclusion} \label{6}
The paper discussed the design issues of a turbo decoder on a TTA processor. The reference turbo decoder design has been simulated on a Matlab link level simulator. The design is then converted and optimized in C language and mapped on the TTA processor using TCE tool. The design shows the promise of the possibility of designing several decoding techniques on a single TTA processor.  As the turbo decoding algorithm is very complex, it is very difficult to achieve the LTE target throughput even with pure hardware designs. The parallel multi-core turbo decoder is the natural choice for the next generation wireless systems. The target throughput could also be reached by multi-core TTA processor. The flexibility gained from that processor could provide very interesting results and would be a fruitful direction for future research.

\section{Acknowledgement} \label{con}
This research was supported by the Finnish Funding Agency for Technology and Innovation (Tekes), Renesas Mobile Europe, Nokia Siemens Networks, Elektrobit and Xilinx. 
Special thanks are due to Dr. Perttu Salmela from Tampere University of Technology for sharing his insights on programmable turbo decoder implementations.

{}

\end{document}